\documentclass[aps,twocolumn,pra]{revtex4-1}

\usepackage{newlfont}
\usepackage{color}         
\usepackage{amssymb}
\usepackage{amsfonts}
\usepackage{amsmath}
\usepackage{wasysym}
\usepackage{graphicx}
\usepackage{epsfig}
\usepackage{amsthm}
\usepackage{bm}
\usepackage{mathrsfs}
\usepackage{epstopdf}
\usepackage{theorem}

\usepackage[colorlinks=true, citecolor=blue, urlcolor=blue ]{hyperref}

\begin{document}

\title{Activation of  Nonmonogamous  Multipartite Quantum States}

\author{Saptarshi Roy, Tamoghna Das, Asutosh Kumar, Aditi Sen(De), Ujjwal Sen}

\affiliation{Harish-Chandra Research Institute, Chhatnag Road, Jhunsi, Allahabad 211 019, India, and\\
Homi Bhaba National Institute, Anushaktinagar, Mumbai - 400094, India}

\begin{abstract}


The monogamy relation for quantum correlations  is not satisfied by all measures for all multiparty quantum states. We prove that 
an arbitrary quantum state which is nonmonogamous for negativity will become monogamous  if a finite number of copies of the same state is provided. We refer this as activation of nonmonogamous states. We also show that multiple copies of a state satisfy monogamy for negativity if it does so for a single copy.  The results are true for all quantum states of an arbitrary number of parties. Moreover, we find that two different three-qubit pure states which individually violate monogamy relation for negativity, taken together can satisfy the three-party monogamy relation.   This holds for almost all three-qubit pure states while it is true for all three-qubit pure states when a four-party monogamy relation is used to check for their activation. We finally connect monogamy of negativity with genuine multipartite  entanglement.

\end{abstract}

\maketitle

\section{Introduction}

Quantum mechanical postulates of linearity and  unitarity put restrictions on  tasks like  cloning, broadcasting and deleting   \cite{nocloning, nobroadcast, nodeleting, nobitcom, nosplit} which can be executed efficiently in a classical world. However, these no-go theorems  turn out   to be crucial in several quantum information  protocols, including secure quantum communication \cite{qkdRMP} and quantum teleportation \cite{teleportation}.

Constraints in a many-body scenario include the existence of bound entangled states \cite{boundent} and unconvertible state pairs by local quantum operations and classical communication (LOCC) \cite{Nielsenandothers}.  Bound entangled states are ones that require entanglement for their creation implying a nonvanishing entanglement cost \cite{boundent}, although after preparation, 
it is impossible to distil \cite{distill} its entanglement content. On the other hand, it was 
shown  by using the mazorization criterion that there are pairs of bipartite pure quantum states 
that cannot be transformed into each other by LOCC \cite{Nielsenandothers}. Existence of these states seems to limit the  protocols that can be performed by using them. However,  this is  not true in general. It was found that although the bound entangled states are not useful for 
quantum teleportation \cite{Boundcrypto}, these states can be useful in increasing the teleportation fidelity of  distillable states \cite{boundactivation} 
(see also \cite{Boundcrypto}).  It was also shown that a pair of bipartite quantum states can be distillable even though the individual states are respectively bound entangled with 
positive  partial transpose and conjectured bound entangled with negative partial transpose \cite{Shor-Smolin-Terhal}. In a different spirit, a locally unconvertible bipartite pure quantum state pair may become convertible with the help of another    entangled state, a process known as catalysis \cite{catalysis}.

In the multiparty domain, there appears the restriction of monogamy so that entanglement can not be shared arbitrarily between different partners. Monogamy of entanglement puts limitations on sharing of entanglement among different subsystems of a multipartite state which is not the case for the distribution of classical correlation \cite{crypto_monogamy, ent_monogamy, CKW}. Monogamy of entanglement has practical applications in a variety of areas including quantum cryptography \cite{crypto_monogamy}. In this sense,  monogamous states  can be thought of as  a useful resource in quantum communication between several parties. 
However, there exist multiparty entangled states which are nonmonogamous under the known measures of quantum correlation like 
negativity \cite{negativity, negativitymonogamy}, entanglement of formation \cite{EoF, ent_monogamy}, quantum discord \cite{discord, discomono}, and quantum work-deficit \cite{wd, Salini}. 

\begin{figure*}[t]
\begin{center}
    \includegraphics[width=2\columnwidth,keepaspectratio,angle=0]{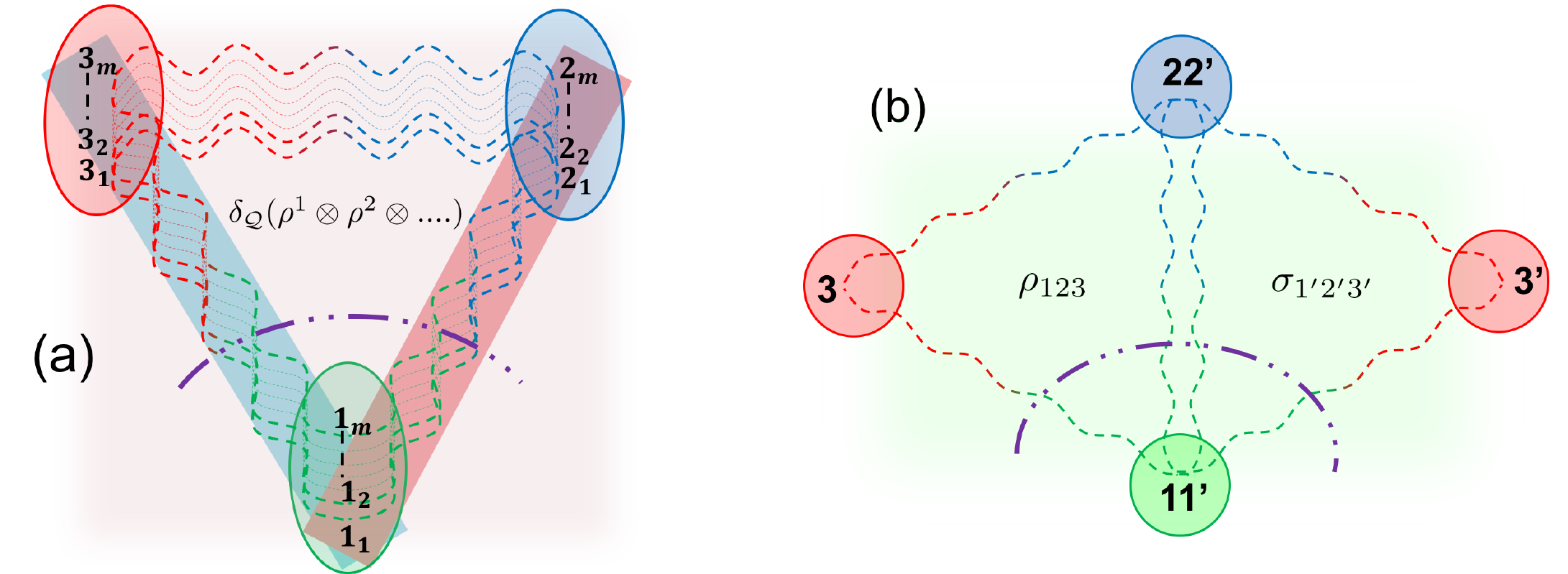}
 \end{center}
\caption{(Color online.) The two scenarios of activation. (a) The multicopy case. Schematic diagram of \(\delta_{\cal Q}(\rho^1_{1_12_13_1} \otimes \rho^2_{1_22_23_2} \otimes \ldots)\), where three-party monogamy score is considered. For activation, we consider two scenarios -- (i) when all \(\rho^i\)'s are the same and (ii) when they are different. In both the cases, activation  occurs, when 
\(\delta_{\cal Q}(\rho^1_{1_12_13_1} \otimes \rho^2_{1_22_23_2} \otimes \ldots) \geq 0\) although \(\delta_{\cal Q} (\rho^i) <0\).  (b) A schematic representation of four-party monogamy score of  \(\delta_{\cal Q}(\rho_{123} \otimes \sigma_{1'2'3'})\), with 3 and 3' being located on separate locations. We look for positivity of \(\delta_{\cal Q}\)($\rho_{123}\otimes\sigma_{1'2'3'}$) with \(\delta_{\cal Q} (\rho_{123}) <0\) and \(\delta_{\cal Q} (\sigma_{1'2'3'}) <0\)   }
\label{fig:Schematic}
 \end{figure*}

At this point,  a natural question can be about the conditions for which multiparty states become monogamous with respect to some quantum correlation measure. Specifically, in this paper,   
we address the following question:

\emph{ Given a bipartite quantum correlation measure ${\cal Q}$, if a multiparty quantum state is nonmonogamous under ${\cal Q}$, is it possible to  obtain monogamy with respect to ${\cal Q}$ when finite copies of the same state are available  or with addition of another state which, considered alone, is also nonmonogamous with respect to ${\cal Q}$?}
 
We find that the answer is in the affirmative for a non-additive quantum correlation measure and we term the phenomenon as activation of nonmonogamous states.
In particular, we  prove  that in the case of negativity, if a multiparty state is monogamous, then its multicopy versions, i.e., several copies of the state also remains so.  We then show that   
a nonmonogamous state with respect to negativity becomes monogamous, when several copies of the same state are provided. The result is true for both pure and mixed states for arbitrary number of parties.  From  extensive numerical searches  for  three-qubit pure states, 
 we observe that to get rid of nonmonogamous nature of a state, the states from the  W-class  require much higher number of copies than the states from the Greenberger-Horne-Zeilinger (GHZ)-class \cite{GHZvsW,GHZ,Wstate,Dur_SLOCC}.
We also prove  that if  two different states  individually satisfy the monogamy inequality for negativity, they continue to obey so even jointly. 
Moreover, we observe that almost all  pairs of three-qubit pure states can jointly obey three-party monogamy with respect to  negativity, even when they individually are nonmonogamous. We also 
introduce a different kind of activation protocol of an $N$-party state, where two copies of $N$-party state is 
considered in an $N+1$-party situation, and the monogamy of the state in the $N+1$-party situation is considered. See Fig. 1(b) for an illustration. In this situation, we numerically confirm that all three-qubit nonmonogamous states 
can be activated for negativity.  
Finally, we establish a relation between ``monogamy scores" for  negativity of  single or two copies of three-qubit pure states and the  amount of a genuine multipartite entanglement.

The paper is organized as follows.  In Sec. \ref{monogamyscore}, we introduce the concept of activation of nonmonogamous states after briefly defining the monogamy score for a quantum correlation measure. Negativity and its closed form for multiple copies of arbitrary states are discussed in the next  section (Sec. \ref{negativity}). The main results of this paper are presented in Sec. \ref{Sec:activateasol}. Specifically, 
we derive the effects on monogamy score for negativity with  multiple copies  in Sec. \ref{Subsec:multicopy}, and nonadditivity of monogamy score for negativity is shown in Sec. \ref{Subsec:nonaddtivity}. In Sec. \ref{Sec:connection}, we make a connection between negativity monogamy score and a genuine multipartite entanglement. We present a conclusion in Sec. \ref{conclusion} .



\section{Monogamy score of quantum correlation measures}
\label{monogamyscore}
Suppose $\rho_{12\ldots N}$ is an arbitrary  \(N\)-party quantum state, and let $\cal Q$ be a bipartite entanglement measure. 
The ``monogamy score" \cite{crypto_monogamy,ent_monogamy,CKW, monogamyscore} of  $\cal Q$ for the state $\rho_{12\ldots N}$, with the party 1 as the ``nodal" observer, is defined as
\begin{equation}
\delta_{\cal Q}\equiv \delta_{\cal Q}(\rho_{12\ldots N}) = {\cal Q}_{1:{\text rest}} - \sum_{i = 2}^N {\cal Q}_{1:i},
\end{equation}
where \({\cal Q}_{1:{\text rest}} \equiv {\cal Q}(\rho_{1:{\text rest}})\) and  \( {\cal Q}_{1:i} \equiv {\cal Q}(\rho_{1:i})\). 
One can consider another party as the nodal observer as well. \(\delta_{\cal Q} (\rho_{12\ldots N}) \geq 0\) implies that the  state, \(\rho_{12\ldots N}\), is monogamous with respect to \({\cal Q}\), and otherwise it is nonmonogamous.
And when it is non-negative for all states, then the quantum correlation measure     \({\cal Q}\) is  said to be monogamous. If there exists a state for which ${\cal Q}$ fails to satisfy the monogamy condition, then \({\cal Q}\) is called nonmonogamous. 
 The squares of  concurrence \cite{CKW},  negativity \cite{negativitymonogamy}, and quantum discord 
 \cite{discordsquare} have been shown to satisfy monogamy, although without the power, they violate the relation 
 \cite{discomono, asu-pra91}. Recently, it was shown that in general, given any quantum correlation measure, \({\cal Q}\), for which \(\delta_{\cal Q} < 0\) for some state, it is always possible to find a monotonically increasing function of \({\cal Q}\) satisfying monogamy for that state \cite{Salini}. In this paper, instead of taking functions of \({\cal Q}\), we choose a different path to obtain monogamy.    

\subsection{Activation vs. additivity of quantum correlation}

Let us consider an \(N\)-party state $\rho_{12\ldots N}$ which does not satisfy the monogamy relation for some \({\cal Q}\), i.e.,  \(\delta_{\cal Q}(\rho_{12\ldots N}) <0\). Considering $m$ copies of the same state, the monogamy score is
\begin{eqnarray}
\delta_{\cal Q}^{(m)}\equiv \delta_{\cal Q}(\rho_{12 \ldots N}^{\otimes m})  =  {\cal Q}_{1_11_2\ldots 1_m:{\text rest}} - \sum_{i = 2}^N {\cal Q}_{1_1 1_2\ldots 1_m:i_1 i_2 \ldots i_m}, \nonumber\\
\end{eqnarray}
where the notation \(i_r (r=1, 2, \ldots m)\) denotes the \(i\)-th   party of the  \(r\)-th  copy of the state.  
For a given \({\cal Q}\), we want to know the status of \(\delta_{\cal Q}(\rho_{12 \ldots N}^{\otimes m}) \). 
If we find that \(\delta_{\cal Q}(\rho_{12 \ldots N}^{\otimes m}) \geq 0\), with   \(\delta_{\cal Q}(\rho_{12\ldots N}) <0\), we say that ``activation" of the nonmonogamous 
state, \(\rho_{12 \ldots N}\), has occurred with respect to \({\cal Q}\). See  Fig. \ref{fig:Schematic}(a)  for a schematic representation of the activation.

In another case, we also want to know the status of  \(\delta_{\cal Q}(\rho_{12 \ldots N} \otimes \sigma_{1'2' \ldots N'})\) where \(\rho_{12 \ldots N}\) and \(\sigma_{1'2' \ldots N'}\)  are both $N$-party states. Specifically, we want to see whether for a given \({\cal Q}\) and \(\rho_{12 \ldots N}\) with  \(\delta_{\cal Q}(\rho_{12 \ldots N}) <0\), it is possible to find a \(\sigma_{1'2' \ldots N'}\) having \(\delta_{\cal Q}(\sigma_{1'2' \ldots N'}) <0\) such that  \(\delta_{\cal Q}(\rho_{12 \ldots N} \otimes \sigma_{1'2' \ldots N'}) \equiv {\cal Q}_{11':{\text rest}} - \sum_{i,i' = 2}^{N, N'} {\cal Q}_{11':i i'} \geq 0\). We again call  this situation as activation. 

In the two preceding cases, we have considered \(N\)-party monogamy scores for \(N\)-party quantum states, even when more than a single copy of the state is available. Relaxing this restriction opens up a large number of possibilities, of which we choose only one. We consider an $(N+1)$-party monogamy score,
\(\delta_{\cal Q}(\rho_{12 \ldots N} \otimes \sigma_{1'2' \ldots N'}) \equiv {\cal Q}_{11':{\text rest}} - \sum_{i,i' = 2}^{N-1, N'-1} {\cal Q}_{11':i i'} - {\cal Q}_{11':N} - {\cal Q}_{11':N'}\), and find whether \(\delta_{\cal Q}(\rho_{12 \ldots N} \otimes \sigma_{1'2' \ldots N'}) \),  can be positive when both \(\delta_{\cal Q}(\rho_{12 \ldots N}) <0\) and \(\delta_{\cal Q}(\sigma_{1'2' \ldots N'}) <0\). This situation can also be called activation of nonmonogamous states. See Fig. \ref{fig:Schematic}(b) for a schematic description.

In this respect, it is interesting to note that any quantum correlation measure, \({\cal Q}\)  is said to be additive, if \({\cal Q}(\rho \otimes \sigma) = {\cal Q}(\rho) + {\cal Q} (\sigma)\).
For some measure \({\cal Q}\), if we are able to show that activation is possible for all states, it implies that \(\delta_{\cal Q}\) is non-additive.

One should note that both the above questions remain invalid if the entanglement measures are additive like squashed entanglement \cite{squashent}, relative entropy of entanglement for two qubits \cite{relativeent}, and logarithmic negativity \cite{negativity}.

\section{Negativity}
\label{negativity}

Having  defined the activation of monogamy for arbitrary \({\cal Q}\), in the rest of the paper, we  consider a specific  entanglement measure, namely negativity.  We begin by giving a  the definition of negativity \cite{negativity}. Thereafter, we discuss the closed form of negativity for multiple copies of a given state, and also when a pair of two different states is considered .  

The negativity of a bipartite quantum state $\rho_{12}$ defined on the composite Hilbert space of ${\cal H}_1^{d_1}\otimes {\cal H}_2^{d_2}$, is based on the partial transposition criterion  \cite{partialtranspose}. If a bipartite quantum state is separable, then the partial transposed state,  $\rho_{12}^{T_1}$, with transposition being taken with respect to the first party,  is  positive semi-definite. 
The negativity of $\rho_{12}$ is defined as
\begin{equation}\label{Eq:negativity_def}
{\cal N}_{12} \equiv {\cal N}(\rho_{12}) = \frac{||\rho_{12}^{T_1}||_1 - 1}{2},
\end{equation} 
where $||\rho||_1$ is the trace norm \cite{fidelity_trd}, defined as $||\rho||_1 = {\text tr}(\sqrt{\rho^{\dagger}\rho})$.
The above equation reduces to 
\begin{equation}\label{Eq:negativity_def_mod}
{\cal N}(\rho_{12}) = \sum_j | \lambda_j^n|,
\end{equation}
where $\lambda_j^n$ are the negative eigenvalues of $\rho_{12}^{T_1}$. Since partial transposition does not change the trace of the matrix,  we have
 $\text{tr}(\rho_{12}^{T_1}) = \sum_j \lambda_j^n + \sum_k \lambda_k^p = 1$, where $\lambda_k^p$ are non-negative eigenvalues. The total number of eigenvalues of $\rho_{12}^{T_1}$  is  $d_1d_2$, and we denote them as $\{\lambda_i\}$. The $\lambda_i$'s can be either $\lambda^{n}_{j}$ or $\lambda^{p}_{k}$.

\subsection{Closed form of negativity with multiple copies and a pair of states}

Let us first calculate the negativity of $m$ copies of   $\rho_{12}$, where $m \in \mathbb{N}$. Without loss of generality, we assume that  in all the copies, partial transposition is taken in the first part.  Note that $\big(\rho_{12}^{\otimes m}\big)^{T_{1_11_2\ldots 1_m}} = \big(\rho_{12}^{T_1}\big)^{\otimes m}$. Hence, the eigenvalues of  \(\big(\rho_{12}^{\otimes m}\big)^{T_{1_11_2\ldots 1_m}}\) are $\{\Pi_{j=1}^m \lambda_{i_j}\}$, Here $\rho_{12}^{ \otimes m}$ is the shorthand for $\rho_{1_12_1} \otimes \rho_{1_22_2} \otimes \ldots \otimes \rho_{1_m2_m}$.
Therefore, we get
\begin{eqnarray}
\label{Eq:neg_Ncopy}
{\cal N}(\rho_{12}^{\otimes m}) &=& \sum_{r = odd} \sum_{\substack{j_1\ldots j_r \\ k_{r+1}\ldots k_m}}  \binom{m}{r} |\lambda_{j_1}^n|\ldots |\lambda_{j_r}^n| \lambda^p_{k_{r+1}}\ldots \lambda^p_{k_{m}} \nonumber \\
&=& \sum_{r = odd} \binom{m}{r} {\cal N}_{12}^r \big(1+{\cal N}_{12}\big )^{m-r} \nonumber \\
&=& \frac 12 \Big[\big(1 + 2{\cal N}_{12}\big)^m -1 \Big].
\end{eqnarray}
Similar calculation also leads to 
\begin{eqnarray}
\label{Eq:rhotimessigma}
{\cal N}(\rho_{12}\otimes \sigma_{1'2'}) &=& \sum_{j, k} (|\lambda_{j_\rho}^n| \lambda_{k_\sigma}^p + |\lambda_{j_\sigma}^n| \lambda_{k_\rho}^p) \nonumber\\
&=& {\cal N}^{\rho} ( 1 + {\cal N}^{\sigma}) +  {\cal N}^{\sigma} ( 1 + {\cal N}^{\rho}), 
\end{eqnarray}
where \( \lambda_{j_\rho}\) and \(\lambda_{j_\sigma}\) denote  the eigenvalues of \(\rho\) and \(\sigma\) respectively, and the superscripts $p$ and $n$ respectively indicate when the eigenvalue is non-negative and negative. Here \({\cal N}^{\rho}\) represents \({\cal N}(\rho_{12})\). Similarly,  \({\cal N}^{\sigma} \equiv {\cal N}(\sigma_{12})\).

\section{Activating  nonmonogamous states with respect to negativity}
\label{Sec:activateasol}
In this section, we prove the main results. The first subsection is devoted to the situation where many copies of a state is considered, and a single copy of which violates monogamy for negativity. Next, we discuss the status of  monogamy score of negativity for a pair of states which are separately nonmonogamous. 

\subsection{Effect of multiple copies on monogamy}
\label{Subsec:multicopy}

Before answering the main question, we first ask the following question: If negativity monogamy score is non-negative for a single copy of a state, will it remain so with multiple copies of the same state? The answer is ``yes'' and it holds in arbitrary dimensions and for arbitrary number of parties. The proofs are mainly given for  three-party states, which can be easily generalized to arbitrary number of parties.   

\text{\bf Theorem 1.} 
If  negativity of a tripartite state, \(\rho_{123}\), satisfy monogamy i.e. if $\delta_{\cal N}(\rho_{123}) \geq 0$, then the $m$ copies of the same state  also remains monogamous. 

\textit{Proof :} 
By using Eq. (\ref{Eq:neg_Ncopy}), the monogamy score of negativity
for the m copies of $\rho_{123}$ can be written as
\begin{eqnarray}
\label{Eq:mscore_rho_m}
\delta_{\cal N}(\rho_{123}^{\otimes m}) &=& {\cal N}(\rho_{1:23}^{\otimes m}) - {\cal N}(\rho_{12}^{\otimes m}) - {\cal N}(\rho_{13}^{\otimes m}) \nonumber \\
&=& \frac 12 \big[(1 + 2~{\cal N}_{1:23})^m  - (1 + 2~{\cal N}_{12})^m \nonumber \\ 
&&  \hspace{2cm} - (1 + 2~{\cal N}_{13})^m +1\big].
\end{eqnarray} 
We now use the method of induction to prove $\delta_{\cal N}(\rho_{123}^{\otimes m}) \geq 0$ for all $m$, provided  $\delta_{\cal N}(\rho_{123}) \geq 0$.

For $m= 2$, Eq. (\ref{Eq:mscore_rho_m}) reduces to
 \begin{eqnarray}
\delta_{\cal N}(\rho_{123}^{\otimes 2}) 
&=& 2\big[{\cal N}_{1:23}(1 + {\cal N}_{1:23}) - {\cal N}_{12}(1 + {\cal N}_{12}) \nonumber \\ 
&&  \hspace{3cm} - {\cal N}_{13}(1 + {\cal N}_{13})\big] \nonumber \\
&\geq & 2\big[({\cal N}_{12} + {\cal N}_{13})(1 + {\cal N}_{12} + {\cal N}_{13})  \nonumber \\ 
&& \hspace{0.5cm} - {\cal N}_{12}(1 + {\cal N}_{12}) - {\cal N}_{13}(1 + {\cal N}_{13})\big] \nonumber\\
&=& 4~{\cal N}_{12}{\cal N}_{13} \geq 0, 
\end{eqnarray}
where the inequality follows from  $\delta_{\cal N}(\rho_{123}) \geq 0$.
Thus the theorem is true for $m = 2$.  Let us now assume that  \(\delta_{\cal N}(\rho_{123}^{\otimes m}) \geq 0\) for some $m = n$, i.e., 
\(\delta_{\cal N}(\rho_{123}^{\otimes n}) \geq 0\).
This implies that,
\begin{eqnarray}
\label{Eq:Induction_n_true}
  \ (1 + 2\,{\cal N}_{1:23})^n + 1 &\geq& (1 + 2\,{\cal N}_{12})^n + (1 + 2 \,{\cal N}_{13})^n 
\end{eqnarray}
Then for $m = n+1$, we have
\begin{eqnarray}
&&(1 + 2\,{\cal N}_{123})^{n+1} = (1 + 2\,{\cal N}_{123})^{n} (1 + 2\,{\cal N}_{123}) \nonumber \\
&& \geq \big\{(1 + 2\,{\cal N}_{12})^n + (1 + 2 \,{\cal N}_{13})^n - 1\big\} \times \nonumber \\
&& \hspace{4cm} (1 + 2 \,{\cal N}_{12} + 2 \,{\cal N}_{13}) \nonumber \\
&& = (1 + 2\,{\cal N}_{12})^{n+1} + (1 + 2\,{\cal N}_{13})^{n+1} - 1 \nonumber \\
&& + 2\,{\cal N}_{12}\big\{\underbrace{(1 + 2 \,{\cal N}_{13})^n - 1}_{\geq 0}\big\} + 2\,{\cal N}_{13}\big\{\underbrace{(1 + 2 \,{\cal N}_{12})^n - 1}_{\geq 0}\big\} \nonumber \\
&& \geq (1 + 2\,{\cal N}_{12})^{n+1} + (1 + 2\,{\cal N}_{13})^{n+1} - 1.
\end{eqnarray}
Hence the proof \cite{another_proof}.  \hfill $\blacksquare$

\text{\bf Corollary.} For an $N$-party state,  $\rho_{12 \ldots  N}$, $\delta_{\cal N}(\rho_{12 \ldots N}) \geq 0$ implies $\delta_{\cal N}(\rho_{12 \ldots N}^{\otimes m}) \geq 0$ for some $m\in\mathbb{N}$.\\
\textit{Proof :} The proof is similar to the proof of Theorem 1. Here, we notice  that 
\begin{eqnarray}
\delta_{\cal N}(\rho_{12\ldots N}^{\otimes 2}) \geq 4 \sum_{i=2,  j>i}^{N-1} {\cal N}_{1i} {\cal N}_{1 j},
\end{eqnarray}
which is again a non-negative quantity. Then by using induction, we can have the proof. 
\hfill $\blacksquare$


We now consider activation of nonmonogamous states for negativity. Again, we give the proof for three-qubit mixed states, which can be easily extended to an arbitrary number of parties. 

\textit{\bf Theorem 2:} For a three-qubit mixed state, $\rho_{123}$, if the negativity monogamy score is negative, i.e., if $\delta_{\cal N}(\rho_{123}) < 0$, then 
$\rho_{123}^{\otimes m}$ becomes monogamous for some positive integer $m$. \\ 
\textit{Proof :} We again note that Eq. (\ref{Eq:mscore_rho_m}) can be expressed with the help of Eq. (\ref{Eq:neg_Ncopy}) as  
\begin{eqnarray}
\label{Eq:mscore_rho_m2}
2\delta_{\cal N}(\rho_{123}^{\otimes m}) &=& 2\big[{\cal N}(\rho_{1:23}^{\otimes m}) - {\cal N}(\rho_{12}^{\otimes m}) - {\cal N}(\rho_{13}^{\otimes m})\big] \nonumber \\
&=& \sum_{k=1}^{m} \binom{m}{k}2^k\big({\cal N}_{1:23}^k - {\cal N}_{12}^k-{\cal N}_{13}^k \big).
\end{eqnarray}
Since negativity is a convex function \cite{negativity}, and the squared negativity is monogamous for three- and more qubit pure \cite{negativitymonogamy, asu-pra91}, and mixed \cite{asu-arx1409.8632} quantum states,  we can rewrite Eq. (\ref{Eq:mscore_rho_m2}) as
\begin{eqnarray}
\label{Eq:mscore_rho_m3}
2\delta_{\cal N}(\rho_{123}^{\otimes m}) 
&=& \sum_{k=2}^{m} \binom{m}{k}2^k\big({\cal N}_{1:23}^k - {\cal N}_{12}^k-{\cal N}_{13}^k \big) \nonumber \\
&& + 2m\big({\cal N}_{123} - {\cal N}_{12}-{\cal N}_{13} \big). 
\end{eqnarray}
Now the last term of the above equation is negative,  since  we assume that \(\delta_{\cal N}(\rho_{123}) <0\) for  \(\rho_{123}\) \cite{asu-pra91}, and it is linear with $m$. On the other hand, the first term is polynomial of degree \(k \geq 2\).  It is therefore always possible to choose some positive integer $m \geq 2$ such that the first term of $\delta_{\cal N}(\rho_{123}^{\otimes m})$ is bigger than the last one.  \\
Hence the proof. 
\hfill $\blacksquare$ 

{\textbf Note 1.} The above proof uses the fact that negativity is convex and its squared monogamy score is non-negative  for pure as well as mixed three-qubit states. Since squared negativity satisfies monogamy  for all states with arbitrary number of qubits, Theorem 2 also holds for arbitrary multiqubit states. 

{\textbf Note 2.} The above theorem also proves that negativity monogamy score is not additive. 

\begin{figure}[ht]
\begin{center}
  \includegraphics[width=1\columnwidth,keepaspectratio,angle=0]{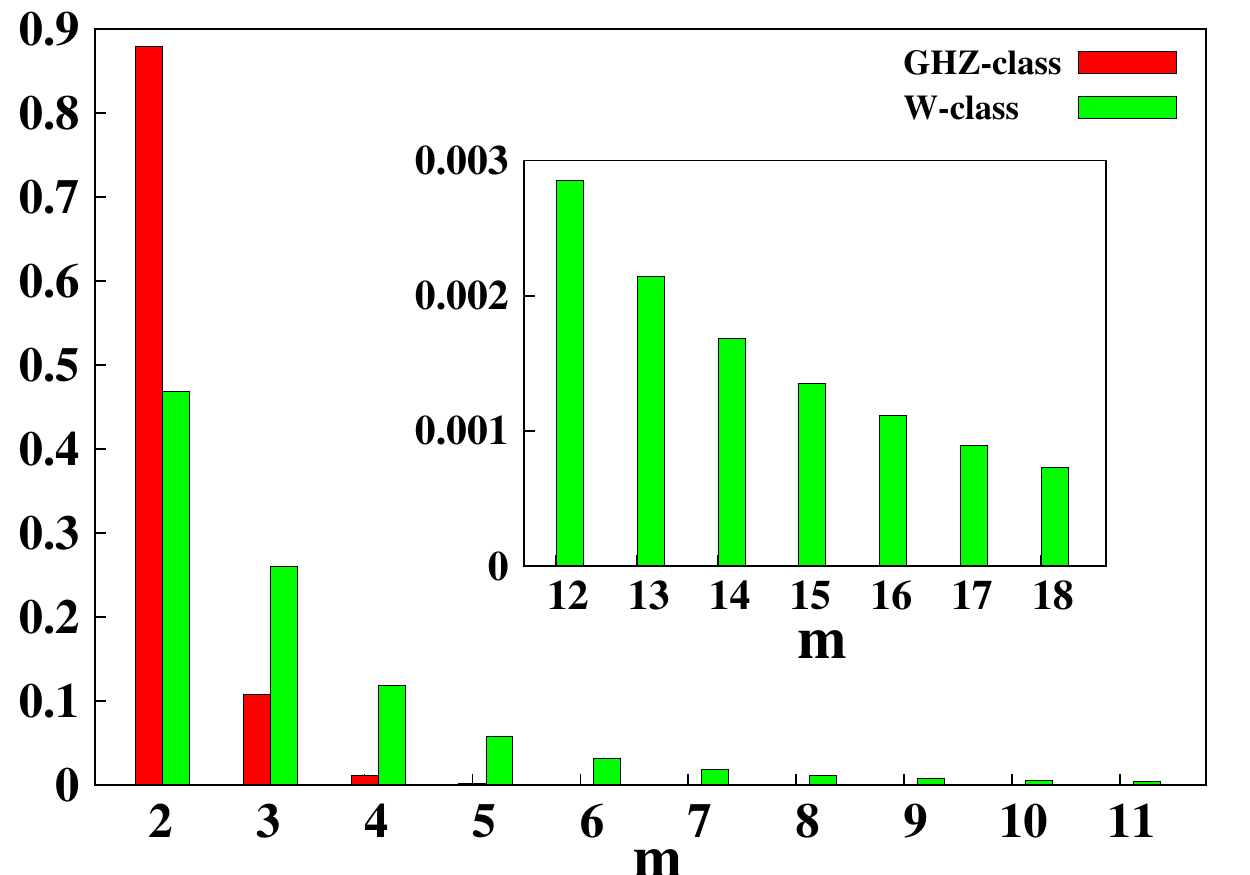}
 \end{center}
\caption{(Color online.) Histogram of probability for obtaining \(\delta_{\cal N}(\rho_{123}^{\otimes m}) \geq 0\),  against the minimum number of copies $m$. We Haar uniformly generate $10^7$ states from both the GHZ- and the 
W-classes. We observe that approximately $88\%$ states  from the GHZ-class, which do not satisfy monogamy when a single copy is provided, become monogamous when  two copies are considered.  For the W-class, the same is $47\%$. 
(Inset) The same plot for only W-class states but for higher $m$. All the axes are dimensionless. }
\label{fig:Histo_neg_m}
 \end{figure}

To visualize the effect of multiple copies on negativity monogamy score, we Haar uniformly generate $10^7$ three-qubit pure states from both the GHZ and the W-classes.   depicts The histogram  in Fig. \ref{fig:Histo_neg_m} depicts the probability  for arbitrary nonmonogamous pure states
becoming activated with the increase of $m$. 
The analysis also reveals that there is a marked difference between the GHZ-class and the  W-class states with respect to their activation. In particular a W-class state which is nonmonogamous with respect to negativity, generally requires a 
higher value of $m$ for satisfying the monogamy relation as compared to a GHZ-class state.   For example, we find that 
there are only few random states from the  GHZ-class, which require more than 10 copies to obtain monogamy for negativity. However, this is not the case for the W-class states as shown in the inset of Fig.  \ref{fig:Histo_neg_m}.

\begin{figure}[ht]
\begin{center}
  \includegraphics[width=1\columnwidth,keepaspectratio,angle=0]{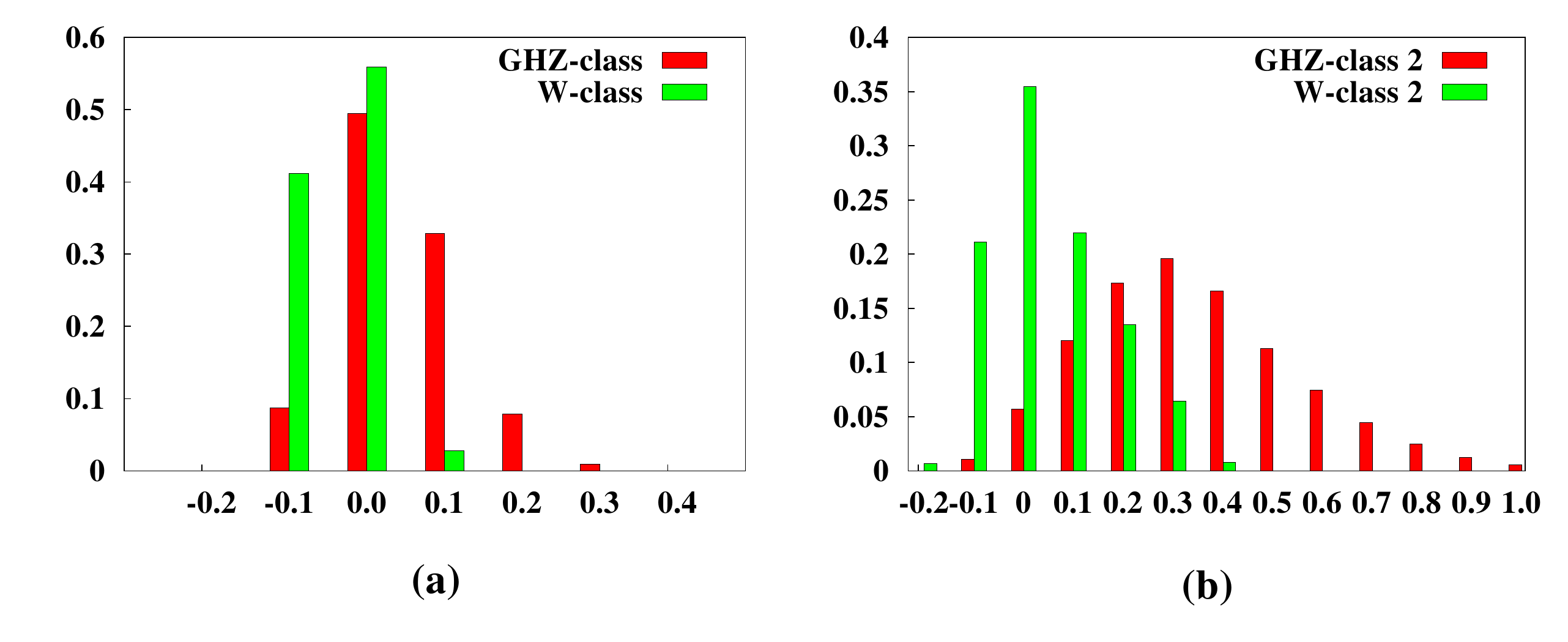}
 \end{center}
\caption{(Color online.) (a) Distribution of negativity monogamy score of three-qubit pure states, \(|\psi\rangle\). 
On the horizontal axis, the value of the tick (\(x_i\)) indicates that the corresponding probability columns are for \(x_i \leq \delta_{\cal N} (\psi) < x_{i+1} \), where \(x_{i+1}\) denotes the value of the next tick. 
 A higher number of states from the GHZ class satisfy monogamy compared to the W-class states, confirming the results in Fig. \ref{fig:Histo_neg_m}. (b) Similar distribution is depicted for \(|\psi\rangle^{\otimes 2}\). Other notations are the same.  We see that almost all states from the GHZ-class becomes monogamous when two copies are considered. The abscissa is in ebits while the ordinate is dimensionless. 
}
\label{fig:Histo_neg_m_2copy}
 \end{figure}

In Fig. \ref{fig:Histo_neg_m_2copy}, we compare the distribution of negativity monogamy scores of \(|\psi_{123}\rangle\) and \(|\psi_{123}\rangle^{\otimes 2}\), when we Haar uniformly generate three-qubit pure states. 
We observe that  almost all states from the GHZ-class become monogamous when two copies are considered, while this is not the case for the states from the W-class. 
The difference in the behavior of the negativity monogamy scores of \(|\psi_{123}\rangle\), and \(|\psi_{123}\rangle^{\otimes 2}\) is analysed in the succeeding section from the perspective of genuine multiparty entanglement. 
 
\subsection{Nonadditivity of monogamy score for negativity}
\label{Subsec:nonaddtivity}
In this section, we choose two tripartite states, \(\rho_{123}\) and \(\sigma_{1'2'3'}\) such that 
\begin{eqnarray}\label{Eq:Non-mono_rho_sigma}
{\cal N}(\rho_{1:23}) &<& {\cal N}(\rho_{12}) +{\cal N}(\rho_{13}), \nonumber \\
{\cal N}(\sigma_{1':2'3'}) &<& {\cal N}(\sigma_{1'2'}) +{\cal N}(\sigma_{1'3'}). 
\end{eqnarray}
We are interested in the sign of 
\begin{eqnarray}
\label{Eq:Non-mono_rho_tensor_sigma}
&& \delta_{\cal N}(\rho_{123}\otimes\sigma_{1'2'3'})  \nonumber \\ 
&& = {\cal N}(\rho_{1:23}\otimes \sigma_{1':2'3'}) - {\cal N}(\rho_{12}\otimes\sigma_{1'2'}) - {\cal N}(\rho_{13}\otimes\sigma_{1'3'}). \nonumber \\ 
\end{eqnarray}
In particular, corresponding to each $\rho_{123}$, our task is to find out $\sigma_{1'2'3'}$ so that 
both relations in Eqs. (\ref{Eq:Non-mono_rho_sigma}) are satisfied, where $\delta_{\cal N}(\rho_{123}\otimes\sigma_{1'2'3'}) \geq 0$.
By using Eq. (\ref{Eq:rhotimessigma}), we obtain that the above relations hold,  only when the expression given by
\begin{eqnarray}
\label{eq:conditionrhosigmathree}
2{\cal N}_{1:23}^{\rho}{\cal N}_{1':2'3'}^{\sigma} &-& \big(2{\cal N}_{12}^{\rho}{\cal N}_{1'2'}^{\sigma} + 2{\cal N}_{13}^{\rho}{\cal N}_{1'3'}^{\sigma} \big) \nonumber \\
  &-& \big( |\delta_{\cal N}(\rho_{1:23})| + |\delta_{\cal N}(\sigma_{1':2'3'})| \big) \geq 0.
\end{eqnarray}
In this case, we numerically simulate  $10^7$  three-qubit pure states from the GHZ-class. We find that it is always possible to find a pure state $\sigma = |\phi\rangle\langle \phi|$ from the GHZ-class which satisfies both the relations 
in  (\ref{Eq:Non-mono_rho_sigma}), and $\delta_{\cal N}(\rho_{123}\otimes\sigma_{1'2'3'}) \geq 0$. 
 Among Haar uniformly generated GHZ-class states, we find that there are only $8.8\%$ states, for which  $\delta_{\cal N} <0$.  

On the other hand,  a large number of states from the W-class show nonmonogamous nature for ${\cal N} $. In particular,  within $10^7$ randomly generated states of the W-class,  $43.3\%$ states are nonmonogamous. Numerical simulations show that in this case, there are less than $1\%$ states which do not satisfy monogamy relation for negativity even with the help of \(\sigma\), chosen from the W-class as well as from the GHZ-class. 

With these observations, we can conclude that almost all three-qubit pure states satisfy negativity monogamy relation with the help of another nonmonogamous state from the same class.    It also confirms the non-additivity of negativity monogamy score.

Let us now consider another activation scenario, given in Fig. \ref{fig:Schematic}(b). We again choose two tripartite states which satisfy the relations in (\ref{Eq:Non-mono_rho_sigma}). We now find the condition  
when the four-party monogamy score becomes non-negative, i.e., when
\begin{eqnarray}
\label{eq:condirhosigmafour}
&& \delta_{\cal N}(\rho_{123}\otimes\sigma_{1'2'3'})  \nonumber \\ 
&& = {\cal N}(\rho_{1:23}\otimes \sigma_{1':2'3'}) - {\cal N}(\rho_{12})\nonumber \\
&& - {\cal N}(\rho_{13}\otimes\sigma_{1'3'}) - {\cal N}(\sigma_{1'2'}) \geq 0. 
\end{eqnarray}
The above condition reduces to 
\begin{eqnarray}
\label{eq:conditionrhosigmafourreduce}
 2{\cal N}_{1:23}^{\rho}{\cal N}_{1':2'3'}^{\sigma} - ( 2{\cal N}_{13}^{\rho} {\cal N}_{1'3'}^{\sigma} + |\delta_{\cal N}(\rho_{1:23})| + |\delta_{\cal N}(\sigma_{1':2'3'})|) \geq 0. \nonumber\\
\end{eqnarray}
This is similar to the one obtained for three-party monogamy score, given in  (\ref{eq:conditionrhosigmathree}). However, comparing both the relations, one should note that  (\ref{eq:conditionrhosigmathree}) contains one extra term which is absent in  (\ref{eq:conditionrhosigmafourreduce}). Therefore, all the states from the GHZ-class which satisfy three-party negativity monogamy relation,  continue to obey four-party monogamy condition.  
It will be interesting to find whether the $1\%$ states from the W-class which cannot  be activated by using three-party negativity monogamy score,  can satisfy the  four-party monogamy relation. Numerically, we find that it is indeed the case. 



\section{Relation of negativity monogamy score with genuine multiparty entanglement measures}
\label{Sec:connection}

\begin{figure}[t]
\begin{center}
  \includegraphics[width=1\columnwidth,keepaspectratio,angle=0]{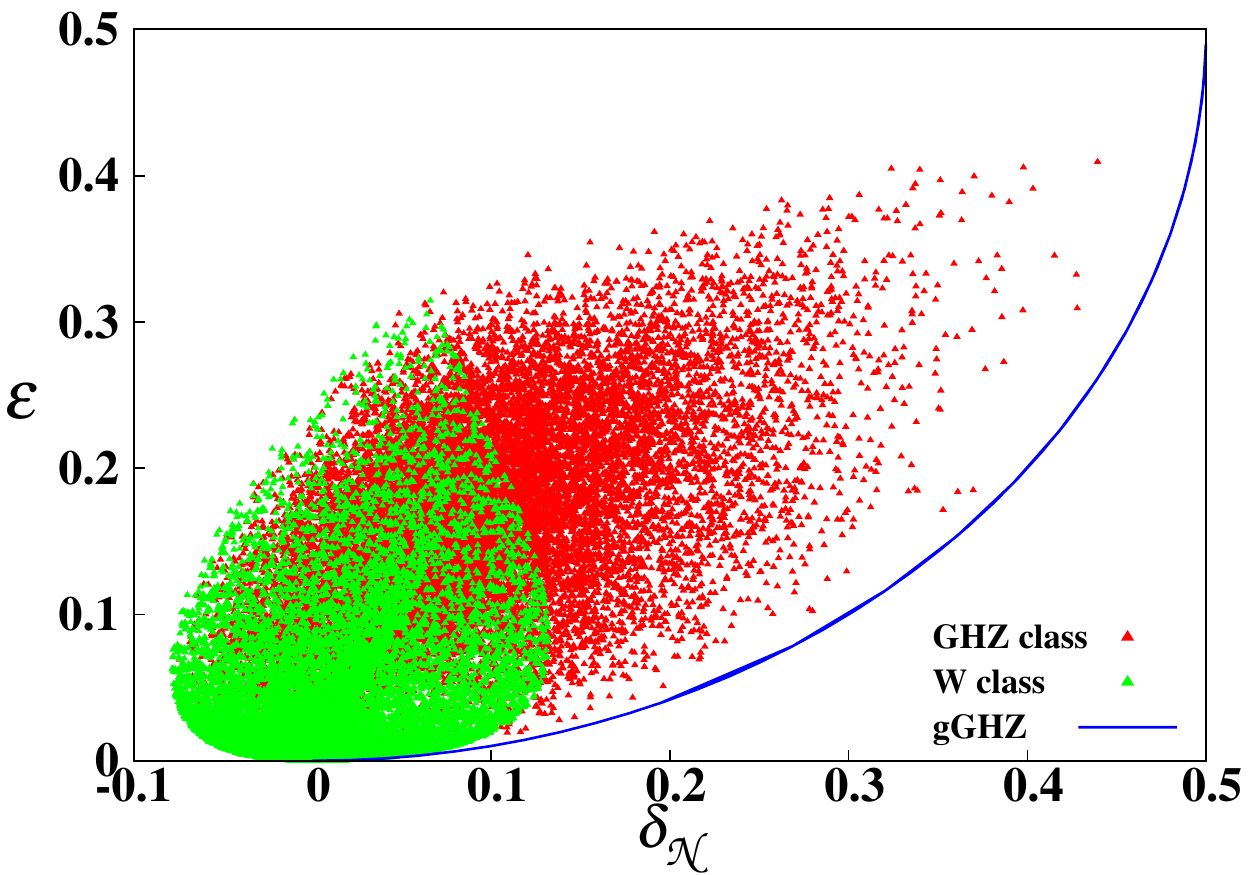}
 \end{center}
\caption{(Color online.) Negativity monogamy score vs.  GGM. We Haar uniformly generate $10^7$ arbitrary three-qubit pure states from the  GHZ (red dots) and the W-classes (green dots). $\delta_{\cal N}$ is plotted along the abscissa while the ${\cal E}$ is plotted along the ordinate. The blue line corresponds to the generalized GHZ state. The horizontal axis is in ebits, where the vertical axis is dimensionless.  }
\label{fig:delta_N_GGM_GHZclass}
 \end{figure}

We now ask whether activation of a state is related to it's multipartite entanglement content. 
 First, we establish  a relation between  $\delta_{\cal N}(|\psi\rangle)$ and  genuine multiparty entanglement measure,  with the later being quantified by the generalized geometric measure (GGM) \cite{GGM} for three-qubit pure states.  
 We then try to build a connection \( \delta_{\cal N}(|\psi\rangle^{\otimes 2})\) and the  GGM of $|\psi\rangle$ of arbitrary $|\psi\rangle$. A multiparty pure quantum state is called  genuinely multiparty entangled if it is entangled in all bipartitions. The generalized geometric measure of an arbitrary state is the minimum distance of the given state from the nongenuinely mutiparty entangled states, and can be simplified for a three-party quantum state in arbitrary dimensions as 
\begin{equation}
{\cal E} (|\psi_{123}\rangle) = 1 - \max \{\lambda_1^{\max}, \lambda_2^{\max}, \lambda_3^{\max}\},
\end{equation}
where \(\lambda_i^{\max} (i=1, 2,3)\)  denotes the maximum eigenvalue of the single-site reduced density matrices, \(\rho_{i} (i=1,2,3)\),  of \( |\psi_{123}\rangle\). 
In this respect, we consider  the generalized GHZ state \cite{GHZ}, given by $|gGHZ\rangle = \sqrt{\alpha}|000\rangle + \sqrt{1-\alpha}e^{i\phi}|111\rangle$ which plays an important role to obtain the connection. In particular, we have the following Proposition:

 \begin{figure}[ht]
\begin{center}
  \includegraphics[width=1\columnwidth,keepaspectratio,angle=0]{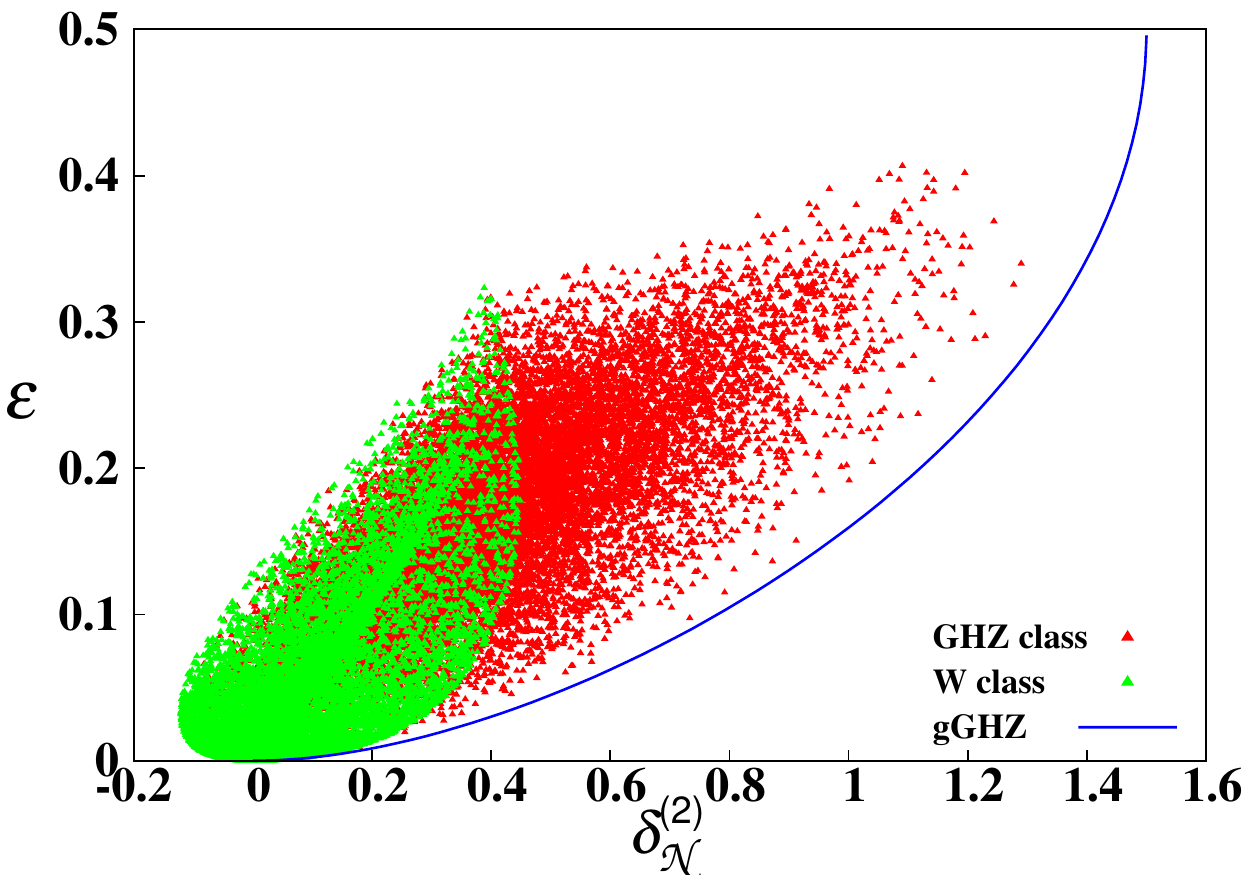}
 \end{center}
\caption{(Color online.) Negativity monogamy score of two copies of randomly generated states against  GGM. All  notations are same as in Fig. \ref{fig:delta_N_GGM_GHZclass}. }
\label{fig:delta_N2_GGM1_GHZclass}
 \end{figure}

\textit{\bf Proposition. } The  GGM of an arbitrary three qubit pure state is bounded below by the GGM of the generalized three qubit GHZ state, whenever $\delta_{\cal N}(|gGHZ\rangle) = \delta_{\cal N}(|\psi\rangle)$ and when the nodal observer gives a maximum eigenvalue involved in  GGM.


\textit{Proof:} The negativity monogamy scores with the first party as the nodal observer, of the  generalized GHZ state  and  an arbitrary three-qubit pure state, $|\psi\rangle$, are respectively given by
\begin{equation}\label{Eq:Neg_mono_gGHZ}
\delta_{\cal N}(|gGHZ\rangle) = \sqrt{\alpha(1-\alpha)},
\end{equation}
\begin{equation}\label{Eq:Neg_mono_psi}
\delta_{\cal N}(|\psi\rangle) = \sqrt{\lambda_1(1-\lambda_1)} - {\cal N}_{12} - {\cal N}_{13} \leq  \sqrt{\lambda_1(1-\lambda_1)} ,
\end{equation}
where \(\lambda_1 \geq 1/2\) is the largest eigenvalue of the local density matrix \(\rho_1\), of \(|\psi\rangle\). Using  (\ref{Eq:Neg_mono_gGHZ}) and (\ref{Eq:Neg_mono_psi}), we get \(\alpha \geq \lambda_1\). Now the GGM of \(|\psi\rangle\) is \(1 - \max\{\lambda_1,\lambda_2, \lambda_3\}\), where \(\lambda_i\) s are the maximal eigenvalues of the local density matrices of $|\psi\rangle$, while ${\cal E}(|gGHZ\rangle) = 1 - \alpha$. If $\lambda_1 \geq \lambda_2, \lambda_3 \geq 1/2$, we have ${\cal E}(|gGHZ\rangle) \leq {\cal E}(|\psi\rangle)\), and 
hence the proof. \hfill $\blacksquare$ 

Numerical simulations indicate that the lower bound is true irrespective of whether the maximum eigenvalue is obtained from the nodal observer or not.  The numerical analysis is performed by generating $10^7$  random three-qubit pure states, Haar uniformly, from both  GHZ and  W-classes (see Fig. \ref{fig:delta_N_GGM_GHZclass}).

Similar analysis can also be carried out between \(\delta_{\cal N}(|\psi\rangle^{\otimes 2})\)  and  \( {\cal E}(|\psi \rangle)\). In this situation,  by using Eq. (\ref{Eq:neg_Ncopy}), it can be proven in a similar fashion  that whenever $\delta_{\cal N}(|gGHZ\rangle^{\otimes2} ) = \delta_{\cal N}(|\psi\rangle^{\otimes 2})$, we have
${\cal E}(|\psi\rangle) \geq {\cal E}(|gGHZ\rangle)$, for arbitrary pure three qubit pure states $|\psi\rangle$, assuming again that the maximum eigenvalue is obtained from the nodal observer.  If we once more lift the restriction of maximum eigenvalue in GGM appearing from the nodal observer,  we see that the \(|gGHZ\rangle\) state  still gives the lower boundary, as seen from Fig. \ref{fig:delta_N2_GGM1_GHZclass}. Moreover, we notice that the states from the W-class which possess low value of GGM and high negative value of \(\delta_{\cal N} (|\psi\rangle)\)  require more than two copies to satisfy the monogamy relation for negativity. Pure states with high  genuine multipartite entanglement which violate monogamy for negativity, are seemingly easier to be activated than the states with low genuine multipartite entanglement.

\section{Conclusion}
\label{conclusion}

Summarizing, we have introduced the concept of activation of nonmonogamous  multiparty states for a non-additive quantum correlation measure.  Our aim was to obtain monogamous states from nonmonogamous ones by using many copies of the same state.   We proved that an arbitrary state, whether pure or mixed,  which does not satisfy the monogamy relation for negativity, will satisfy monogamy when multiple copies of the same state is provided.  This is true irrespective of the  number of parties and their dimension. On the other hand, we showed that negativity monogamy score of multiple copies of a state remains non-negative if it is so for a single copy. We carried out extensive numerical searches among three-qubit pure states and found that nonmonogamous W-class states, generally require a  higher number of copies to satisfy monogamy compared to the states from the GHZ-class. We also analysed the status of monogamy for pairs of states that are individually nonmonogamous. We found that  almost all pairs of states become monogamous, even when they individually are not. Finally, we provided a relation between the negativity monogamy score  and a genuine multipartite entanglement measure for three-qubit pure states. 


\begin{acknowledgments}
We thank Titas Chanda for discussions. 
\end{acknowledgments}





\begin{thebibliography}{19}
\bibitem{another_proof} We now note that the above theorem can also be proved by using binomial expansion.
Eq. (\ref{Eq:mscore_rho_m}) can be expressed as
\begin{eqnarray*}
\hspace{2em}
2\delta_{\cal N}(\rho_{123}^{\otimes m}) &=& 2\big[{\cal N}(\rho_{123}^{\otimes m}) - {\cal N}(\rho_{12}^{\otimes m}) - {\cal N}(\rho_{13}^{\otimes m})\big] \nonumber \\
&\geq & \big[(1 + 2~{\cal N}_{12}) + 2~{\cal N}_{13}\big]^m - (1 + 2~{\cal N}_{12})^m \nonumber \\
&& - (1 + 2~{\cal N}_{13})^m +1 \nonumber \\
&=& \sum_{k=1}^{m-1} \binom{m}{k}~\big(2{\cal N}_{12}\big)^k 
 \big[(1 + 2~{\cal N}_{13})^{m-k}-1 \big].
\end{eqnarray*}
Clearly, the right hand side of the last line is non-negative for $m \geq 1$ which also proves  Theorem 1.
\end{thebibliography}
\end{document}